\documentclass[twocolumn,preprintnumbers,superscriptaddress,nofootinbib,aps,prd,floatfix,longbibliography]{revtex4-2}
\usepackage{graphicx}
\usepackage{amsmath}
\usepackage{subcaption}
\usepackage{lineno}
\usepackage[normalem]{ulem} 
\usepackage{xcolor}
\usepackage{tabularx}
\usepackage{booktabs} 
\usepackage{caption} 
\usepackage{titlesec} 
\usepackage{lipsum} 
\usepackage{hyperref}
\usepackage{xspace}

\newcommand{\PZ}{\ensuremath{\mathrm{Z}}\xspace}
\newcommand{\as}{\ensuremath{\alpha_\mathrm{S}}\xspace}
\newcommand{\asmz}{\ensuremath{\as (m_\PZ)}\xspace}

\newcommand{\NNLL} {\ensuremath{\text{NLO}+\text{NNLL}}\xspace}

\newcommand{\PYTHIA} {{\textsc{pythia}}\xspace}
\newcommand{\HERWIG} {{\textsc{herwig}}\xspace}
\newcommand{\SHERPA} {{\textsc{sherpa}}\xspace}
\newcommand{\xL} {\ensuremath{x_\text{L}}\xspace}

\newcommand{\pbinv} {\mbox{\ensuremath{\,\text{pb}^{-1}}}\xspace}
\newcommand{\epem} {\ensuremath{e^+e^-}\xspace}
\begin{document}

\title{Extracting $\as$ at future $e^+e^{-}$ Higgs factory with energy correlators}
\author{Zhen Lin}
\email{zhenlin@zju.edu.cn}
\affiliation{Zhejiang Institute of Modern Physics, Department of Physics, Zhejiang University,\\
Hangzhou, Zhejiang 310027, China}
\author{Manqi Ruan}
\email{ruanmq@ihep.ac.cn}
\affiliation{
Institute of High Energy Physics, Chinese Academy of Sciences,\\
19B Yuquan Road, Shijingshan District, Beijing 100049, China
}
\author{Meng Xiao}
\email{mxiao@zju.edu.cn}
\affiliation{Zhejiang Institute of Modern Physics, Department of Physics, Zhejiang University,\\
Hangzhou, Zhejiang 310027, China}
\affiliation{Center for high energy physics in Peking University,\\
Beijing 100871, China
}
\author{Zhen Xu}
\email{zhen.xu@zju.edu.cn}
\affiliation{Zhejiang Institute of Modern Physics, Department of Physics, Zhejiang University,\\
Hangzhou, Zhejiang 310027, China}
\date{\today}

\begin{abstract}
The prospected sensitivity in $\as$ determination using an event shape observable, ratio of energy correlators at future electron-positron colliders is presented. The study focuses on the collinear region which has suffered from large theoretical and hadronization uncertainty in the past. The ratio effectively reduces the impacts of the uncertainties. With the amount of data that future electron-positron colliders could produce in 1 minute (40 \pbinv), a 1--2\% precision of \as could be reached depending on the hadronization uncertainty considered.
\end{abstract}

\maketitle

\section{\label{Intro}Introduction}
The strong coupling constant (\as) is a fundamental parameter in Quantum Chromodynamics (QCD). Its value enters theoretical predictions of strong interactions and its extraction requires experimental inputs. The current determination of \as at the Z boson mass $m_\PZ$ has a relative uncertainty of about 1\% ($0.1180\pm{0.0009}$)~\cite{pdg2022}. 
Compared to other constants that describe the strength of fundamental interaction, e.g. the fine-structure constant which characterizes the strength of the electroweak interaction~\cite{RevModPhys.93.025010}, the precision of $\as$ is worse by several orders of magnitude. The limited precision of \as has become one of the dominant uncertainties in the calculations of the Higgs process~\cite{higgs_production_gluonfusion,LHCHiggsCrossSectionWorkingGroup:2016ypw}. It also affects the calculation of the partial widths of the Higgs boson decay~\cite{Lepage:2014fla,LHCHiggsCrossSectionWorkingGroup:2016ypw,Denner:2011mq} and plays a crucial role in determining quantities associated with the top quark, such as its mass, width, and Yukawa coupling~\cite{Hoang_2020,higgs_production_ttbar}.
This imprecision motivates further improvements to enhance our understanding of QCD. 

Various techniques have been deployed to determine $\as$. At \epem colliders, the analyses of event shapes in hadronic final states~\cite{dissertorietal2009,abbiendietal2011,bethkeetal2009,davisonwebber2009,abbateetal2011,gehrmannetal2013,kardosetal2021,hoangetal2015,luisonietal2021,kardosetal2018} have yielded important results. Event shapes are particularly sensitive to soft and collinear physics, which is crucial for refining the precision of \as extraction and facilitating cross-checks with other methods. However, the particular phase space also introduces higher uncertainties: the calculation suffers from additional resummation uncertainties compared to fixed-order calculations. Moreover, the hadronization corrections in this region are often substantial, increasing the non-perturbative uncertainties as well.
For these reasons, the previous extraction of \as using energy energy correlators (E2C) at LEP explored data in the less collinear region~\cite{kardosetal2018}. However the collinear region has rather substantial statistics, it would be beneficial to fully exploit it.

Recently the ratio between the projected three-point energy correlator (E3C) and E2C have proposed to extract \as in the collinear region~\cite{Chen:2020vvp}. This has been performed by CMS experiment~\cite{CMS:2024mlf} and proved to effectively reduce hadronization uncertainty. In this study, we evaluate the feasibility of applying the method in the collinear region at \epem collider at a center-of-mass energy of $\sqrt{s} = 91.2$ GeV. The expected sensitivity of \as is estimated including detector effects, theoretical and hadronization uncertainties. Experimental uncertainties are found to be minor. We present the expected precision of \asmz at a luminosity of 40 \pbinv, which is similar to the combined luminosity used in previous E2C based determinations~\cite{kardosetal2018}. We exploit different ways to evaluate the hadronization uncertainty. A relative precision of 0.8\% in \asmz is obtained when adopting the traditional approach of evaluating Monte Carlo (MC) hadronization uncertainties through model comparisons. However, with more comprehensive decomposition of the uncertainties into multiple sources - including parameter tuning, parton shower (PS) scales and models, hadronization models, color reconnection and generator differences - the estimated precision is 1.5\%. Taking into account the difference between analytical and MC based predictions, the expected precision becomes 2.1\%.

\section{\label{Energy_correlators}Energy correlators and theoretical predictions}

Energy correlators were designed to study the energy flow in an event or a jet~\cite{Basham:1978bw}. The definition is adapted in different colliders~\cite{Chen:2020vvp}. At \epem collider where the energy scale of hard scattering is fixed, the energy correlators describe the energy-weighted angular distances among all the final particles in an event. The simplest energy correlator is E2C, defined as

\begin{align}
   \mathrm{E2C} &= \frac{d\sigma^{[2]}}{d\xL} = \sum\limits_{i,j}^n\int d\sigma\frac{E_iE_j}{Q^2}\delta(\xL-\frac{1-\cos\chi_{ij}}{2}),
\end{align}

where the $\chi$ is the angular distance between particle $i,j$, and the energy of the two particles are denoted by $E_i, E_j$. The E2C observable was extensively studied at early \epem colliders~\cite{SLD:1994idb,L3:1992btq,OPAL:1993pnw,TOPAZ:1989yod,TASSO:1987mcs,JADE:1984taa,Fernandez:1984db,Wood:1987uf,CELLO:1982rca,PLUTO:1985yzc,DELPHI:2000uri}. A previous study has used E2C measurement to extract \as by comparing it to theoretical calculations at NNLO + NNLL precision~\cite{kardosetal2018}. Due to the large uncertainty in both theoretical prediction and hadronization corrections in the collinear region, only the intermediate region data was used, corresponding to $\chi$ between $60^\circ$ to $160^\circ$. The determined result was $\asmz=0.1175 \pm 0.00018(\mathrm{exp.})\pm0.00102(\mathrm{hadr.})\pm0.00257(\mathrm{ren.})\pm0.00078(\mathrm{res.})$. Even so, the theoretical and hadronization uncertainties are notably larger compared to the experimental ones. 

\begin{figure}[htbp]
  \centering
  \includegraphics[width=0.48\textwidth]{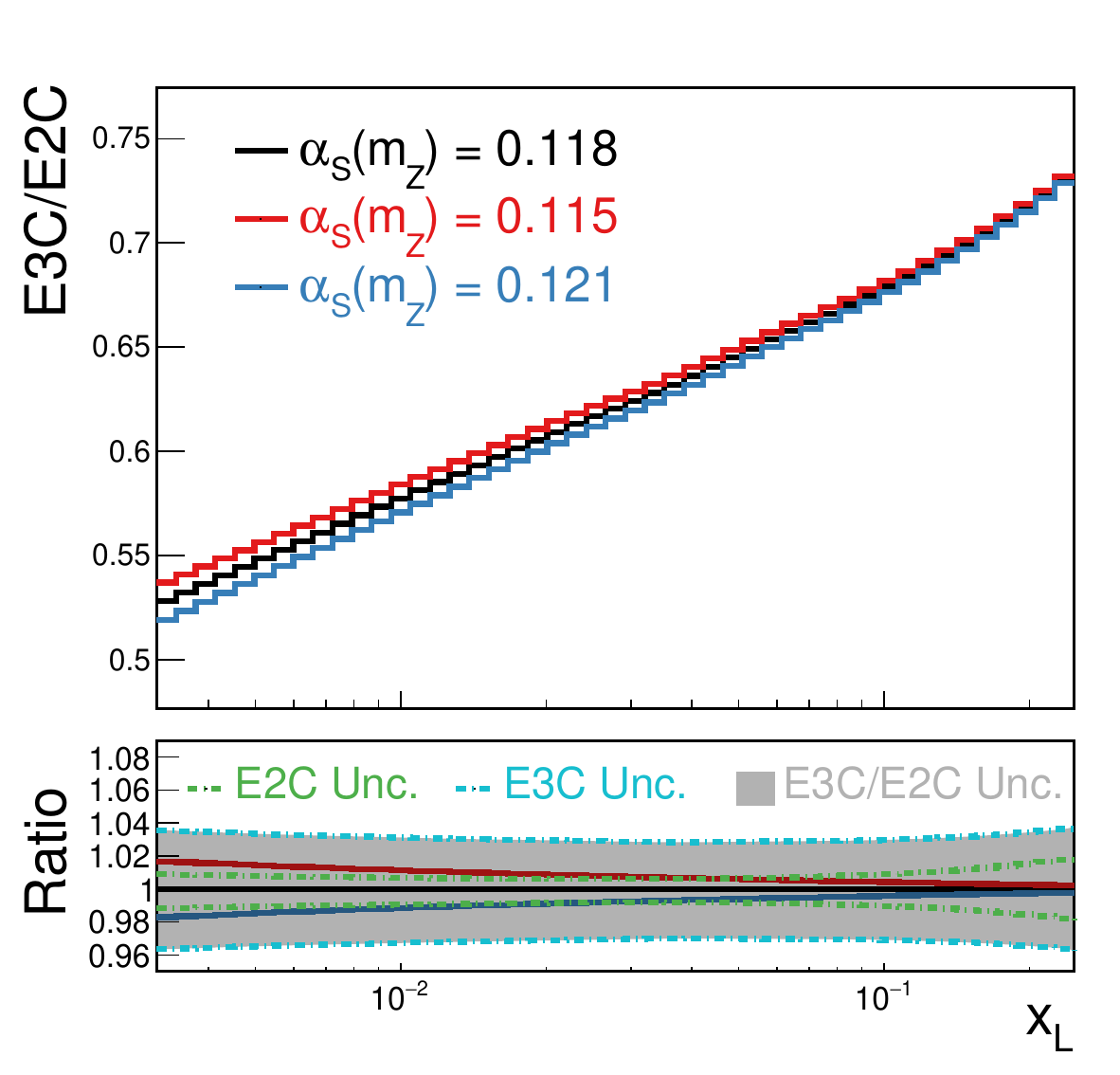} 
  \includegraphics[width=0.48\textwidth]{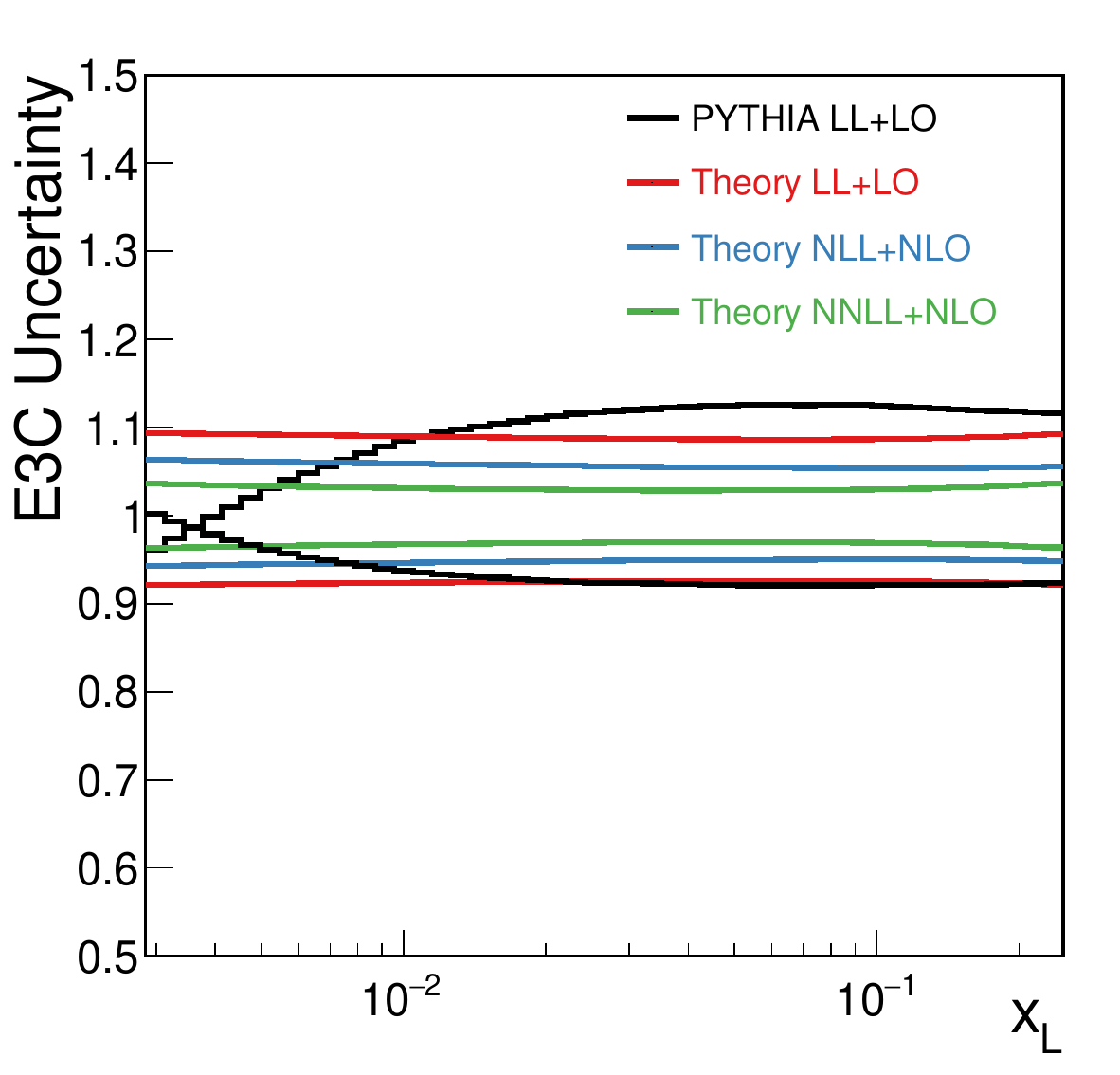} 
  \caption{Top: Theoretical prediction of E3C/E2C for different $\asmz$ values. The lower panel shows the ratio to $\asmz = 0.118$ prediction. Uncertainties of E2C, E2C and their ratio are shown at $\asmz = 0.118$.
 Bottom: E3C uncertainty derived from \PYTHIA simulation and theoretical calculations at different orders.}
  \label{fig:theory_prediction}
\end{figure}

To mitigate such limitation, the ratio of E3C to E2C has been proposed to extract \as in the collinear region~\cite{Chen:2020vvp}. The E3C is an energy correlator that captures the correlation among three particles, defined as 

\begin{equation}
\begin{aligned}
   \mathrm{E3C} &= \frac{d\sigma^{[3]}}{d\xL} = \sum\limits_{i,j,k}^n\int d\sigma\frac{E_iE_jE_k}{Q^3} \\
 &  \delta\left(\xL-
   \max\left(\frac{1-\cos\chi_{ij}}{2},\frac{1-\cos\chi_{ik}}{2},\frac{1-\cos\chi_{jk}}{2}\right)\right).
\end{aligned}
\end{equation}

The ratio of E3C/E2C has minimal non-singular contributions in the collinear limit and allows for safe neglect of higher fixed-order corrections. It also diminishes non-perturbative effects and the associated uncertainties. In order to estimate the sensitivity to \as using the E3C/E2C, we calculate the E3C and E2C at \NNLL precision using the methodology introduced in Ref.~\cite{Chen:2023zlx}, at a center-of-mass energy of $\sqrt{s} = 91.2$ GeV in \epem colliders. It would be interesting to consider the matching to NNLO correction which is available for E2C since Ref.\cite{DelDuca:2016csb} and can be calculated in principle with the methods in Ref.\cite{Gehrmann-DeRidder:2007vsv} and Ref.\cite{DelDuca:2016csb}. However, such a calculation is computationally expensive. We also found that while the matching effect from LO to NLO is 
non-negligible on E2C and E3C respectively, the effect on the ratio is small, especially in the collinear region. Therefore we expect the correction due to further NNLO matching to be minor. 

We consider the $\chi$ angles in the collinear region, from $6^\circ$ to $60^\circ$, corresponding to $\xL=(1-\cos \chi)/2$ in the range of 0.003--0.25. The predicted E3C/E2C distribution at the parton level with different \asmz values are shown in Fig.~\ref{fig:theory_prediction}. A variation of 3\% in \as leads to an approximate 2\% change in the E3C/E2C ratio. 

In the next sections, we will discuss in detail the different types of uncertainties that enter \as determination using E3C/E2C ratio.

\section{\label{Theory}Theoretical scale uncertainties}

For theoretical predictions of E2C and E3C at \NNLL precision, there are two scales that enter the calculation: the hard scale and jet scale. In some earlier studies~\cite{kardosetal2018}, the two scales were varied independently to evaluate the theoretical uncertainties from missing higher order corrections. However, for the calculation used here, resummation is performed by iteratively solving the jet function RG equation to order O($\as^{50}$). Therefore the prediction is regarded as truncated perturbative expansion, where the jet scale and hard scale are related by $\mu_j = \mu_h\sqrt{\xL}$, leaving a single hard scale $\mu_h=\mu$ in the expression. The theoretical uncertainty is therefore evaluated by varying the hard scale $\mu$ by 1/2 and 2. More detailed discussion on the scale uncertainty could be found in Ref.~\cite{Chen:2023zlx}.

The individual theoretical scale uncertainties of E2C and E3C are shown in Fig.~\ref{fig:theory_prediction} ratio panel. As a comparison, we show the scale uncertainty in \PYTHIA 8.306~\cite{Bierlich:2022pfr} simulation in the bottom plot of Fig.~\ref{fig:theory_prediction}, which is at LO+LL, obtained from varying the PS scales of both final state radiation (FSR) and initial state radiation (ISR) by 1/2 and 2. The \PYTHIA PS scale uncertainty is compatible with the theoretical uncertainty at LO+LL precision by varying the hard scale described above. The theoretical scale uncertainties at different perturbative and resummation orders are also shown, where a decrease from 10\% at LO+LL order to 4\% at \NNLL order is seen. The uncertainty in the E3C/E2C ratio, is obtained from a seven-point scale variation method, where the scales of the E2C and E3C are varied independently by a factor of 1/2 and 2, while keeping their ratio between 1/2 and 2. The uncertainty of the ratio is shown as grey band in Fig.~\ref{fig:theory_prediction} top plot. As seen in the ratio panel, taking the ratio does not cancel out scale uncertainties.

\section{\label{Simulation}Non-perturbative effects and uncertainties}
An important aspect of event-shape based \as extraction is the evaluation of non-perturbative corrections. Approaches to derive these corrections can be classified into two general categories: MC based~\cite{kardosetal2018,Verbytskyi:2019zhh,Dissertori:2009qa,Schieck:2012mp,Dissertori:2009ik,OPAL:2011aa,Bethke:2009ehn} and analytical model based~\cite{Tulipant:2017ybb,Lee:2024esz,Davison:2009wzs,Abbate:2010xh,Gehrmann:2012sc,Hoang:2015hka,Bell:2023dqs,Chen:2024nyc,Lee:2024esz}, which do not always yield the same result. In this study, we explore several options in both methods to evaluate the hadronization effects and compare their ultimate impacts on the \as extraction. 

For the analytical model based approach, we used the non-perturbative power correction provided by Ref.~\cite{Lee:2024esz}. It is denoted as $\Omega$ in this paper. For the MC based approaches, the effect is extracted from the distributions predicted by MC at hadron level divided by those at parton level. As there are a variety of models in MC generators, a two-point model uncertainty is usually adopted, where all the aspects of generator setups are varied simultaneously, and the difference, for example, by changing from \PYTHIA to \HERWIG is taken as the uncertainty. Recent guidelines established by Les Houches community~\cite{Andersen:2024czj} have pointed out that a more factorized two-point comparison is needed to assess the different aspects of non-perturbative effects, as has been done by ATLAS collaboration in the jet energy scale uncertainty determination~\cite{ATL-JETM-2022-005,ATL-PHYS-PUB-2022-021}.
Therefore, we try to decompose possible factors in non-perturbative corrections, and evaluate the respective uncertainties either by varying parameters or by utilizing two-point comparisons of settings for a single aspect. The list of factors considered includes:

\newcolumntype{Y}{>{\centering\arraybackslash}X}
\begin{table*}[t]
\centering
\begin{tabularx}{\textwidth}{|Y|Y|Y|Y|Y|Y|}
\hline
\textbf{Generator}        & \textbf{PS model} & \textbf{Hadronization model}  & \textbf{Tune}  & \textbf{PS scale}  & \textbf{CR} \\ \hline
\PYTHIA 8.306            & $p_T$-ordered          & Lund string             & Monash tune  & 1/2, 1, 2 &     MPI,  GM, CS  \\ \hline
\PYTHIA 8.306             & $p_T$-ordered          & Lund string             & Tune:ee = 1--6    & 1     & MPI \\ \hline
\textsc{Herwig} 7.2.2      & Angular-ordered        & Cluster                 & Default     & 1   & Baryonic, plain, statistical     \\ \hline
\textsc{Herwig} 7.2.2      & CS dipole              & Cluster                 & Default      & 1      & Baryonic  \\ \hline
\SHERPA 2.2.12               & CS dipole              & Cluster (AHADIC++)              & Retuned   & 1     & Default   \\ \hline
\SHERPA 2.2.12              & CS dipole              & Lund string             & Default      & 1     & Default    \\ \hline
\end{tabularx}
\caption{ Overview of the generated MC samples. Detailed configurations of the PS, hadronization, tune and CR models are listed. }
\label{tab:mc}
\end{table*}
\begin{figure*}[htbp]
  \centering
    \includegraphics[width=0.48\textwidth]{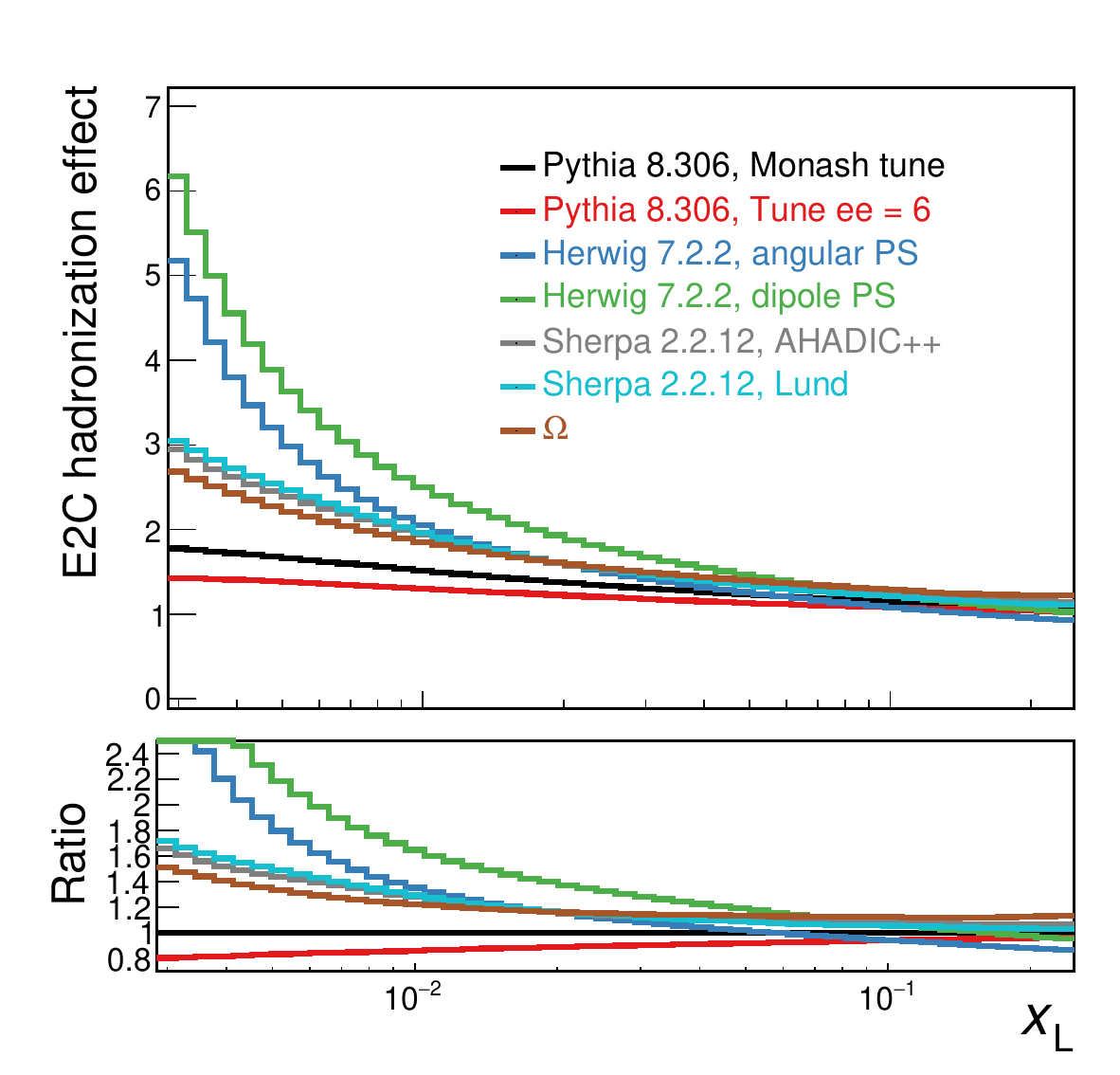}
    \includegraphics[width=0.48\textwidth]{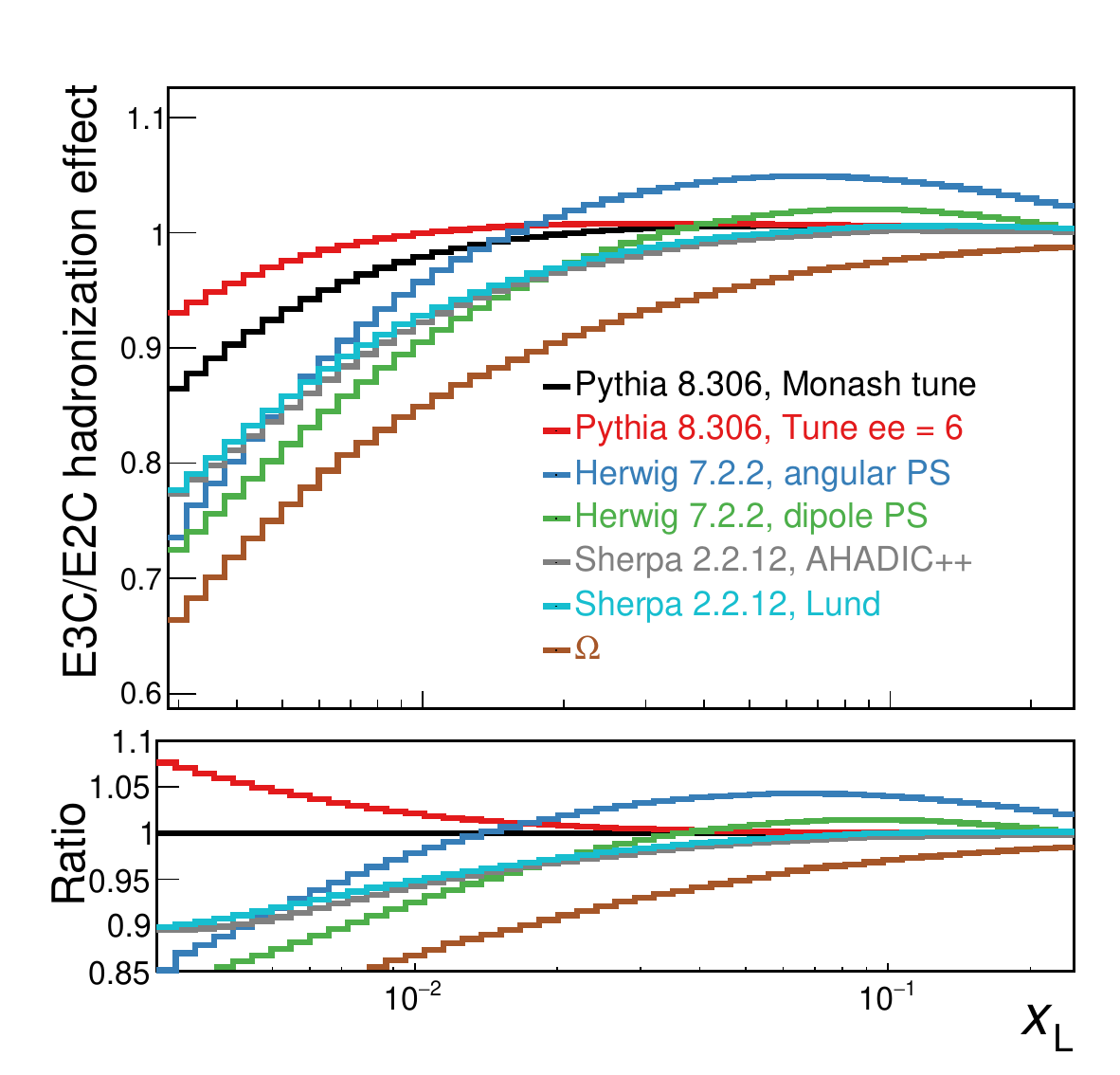}
  \caption{Hadronization correction factors for E2C (left) and E3C/E2C (right), derived from various MC generators and settings.}
\label{fig:combined_hadronization_effects}
\end{figure*}

\begin{itemize}

\item \textbf{PS scale:} PS scale changes the prediction at both parton level and hadron level, and the yielded hadronization correction could be different. We vary the renormalization scales of FSR and ISR independently by factors of 1/2 and 2 from their nominal values, each time a new hadronization correction is calculated, and the uncertainty is obtained from a seven-point scale envelope. This uncertainty is evaluated in \PYTHIA simulation.

\item\textbf{PS model:} Multiple PS models are available in MC generators. The mostly commonly adopted ones include the $p_T$-ordered PS implemented in \PYTHIA, the angular-ordered PS~\cite{Gieseke:2003rz} based on coherent branching implemented in \HERWIG 7.2.2~\cite{Bellm:2015jjp}, and the Catani-Seymour (CS) dipole PS~\cite{Platzer:2009jq,Platzer:2011bc,Schumann:2007mg} implemented in both \HERWIG and \SHERPA 2.2.12~\cite{Gleisberg:2008ta,Sherpa:2019gpd} with customized settings. To isolate the effect of different PS model, we keep the rest of configurations, such as hadronization model and tune parameters the same. This is realized by using \HERWIG simulation, where angular ordered and CS dipole PS are compared while keeping the other settings fixed. 

\item \textbf{Hadronization model:} Two general types of hadronization models are widely used in MC generators: the Lund string model~\cite{ANDERSSON198331,Andersson:1997xwk}, which is deployed in \PYTHIA and \SHERPA, and the cluster-based model~\cite{Webber:1983if} deployed in \HERWIG and \SHERPA. To evaluate the sole effect of hadronization model, two \SHERPA simulations are compared where the CS diople PS is interfaced with either the cluster-based \textsc{AHADIC++} model or the Lund string model.

\item \textbf{Generator:} Even if the general parton shower and hadronization models are the same in different MC generators, the derived hadronization correction could differ due to internal optimizations. This additional uncertainty is determined by comparing \HERWIG with \SHERPA, where both simulations use CS dipole shower and cluster hadronization models.

\item \textbf{Tune:} For hadronization, each generator has its own default set of tune parameters. \PYTHIA provided multiple parameter sets that were derived by different methods. We tested all the seven choices called by Tune:ee method in \PYTHIA, which were tuned to \epem data. The two choices that result in the largest difference in hadronization correction: the default Monash tune optimized for both $e^+e^-$ and $pp$ collisions, and the Tune:ee = 6 are taken as the tune uncertainty. 

\item \textbf{Color reconnection (CR):} Although it is believed that CR has a small impact in \epem colliders, we explicitly tested it by employing different CR models. This is assessed by comparing the default baryonic model with plain and statistical models in \HERWIG. A similar comparison is performed between the default MPI-based model with GM and CS models in \PYTHIA. Both are found to have small effect on the hadronization correction (less than 0.5\%). The slightly more pronounced difference between the baryonic model and the two other models in \HERWIG is taken as the CR uncertainty.
\end{itemize}

Events of $\epem \rightarrow qq$ at $\sqrt{s} = 91.2$ GeV are simulated. This energy has the benefit of a high cross-section of the two-quark final state and negligible background contribution. We employ three event generators: \PYTHIA 8.306, \HERWIG 7.2.2, and \SHERPA 2.2.12, each implementing distinct approaches to parton showering and hadronization to assess the various factors introduced above.

We use \PYTHIA 8.306 with a $p_T$-ordered parton shower, combined with Lund string hadronization model to evaluate the PS scale and tune uncertainties. Monash 2013 tune~\cite{Skands:2014pea} is used as the nominal setting. \HERWIG 7.2.2 offers two complementary parton shower algorithms: the angular-ordered shower and CS dipole shower. Both algorithms are interface with \HERWIG's cluster hadronization model to evaluate the PS model uncertainty. The default tune is used. \SHERPA 2.2.12 with CS dipole PS interfaced with two different hadronization models, its native \textsc{AHADIC++} cluster approach~\cite{Winter:2003tt}, and the Lund string model, to extract the hadronization model uncertainty. The former configuration incorporates specially optimized tune parameters~\cite{Chahal:2022rid,ATL-PHYS-PUB-2022-021} validated against LEP data on jet hadron composition. This tune is found to have negligible impact on the hadronization correction. A summary of all the simulated samples with the key configurations are listed in Tab.~\ref{tab:mc}.

\begin{figure*}[htb]
  \centering
    \includegraphics[width=0.48\textwidth]{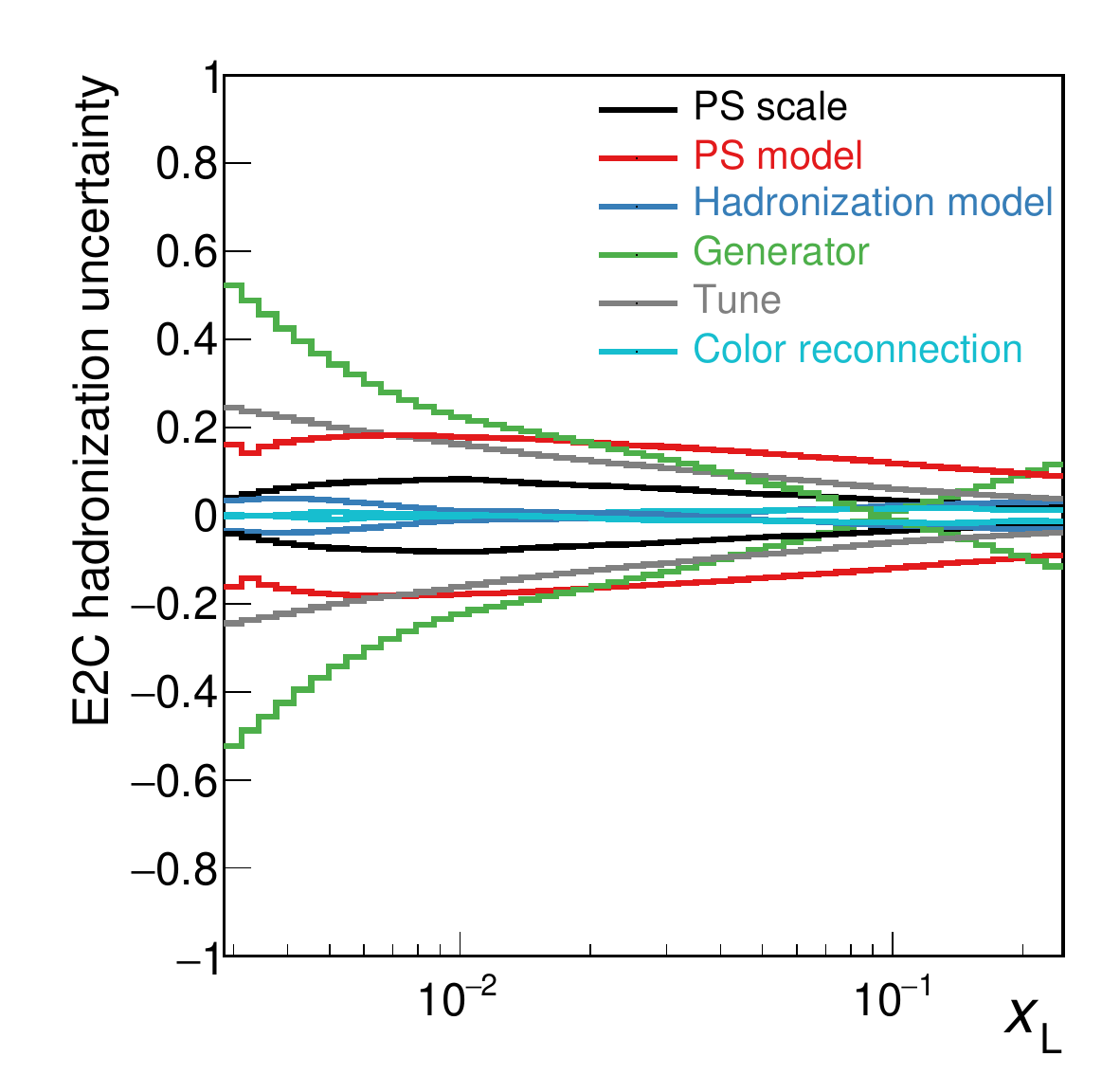}
    \includegraphics[width=0.48\textwidth]{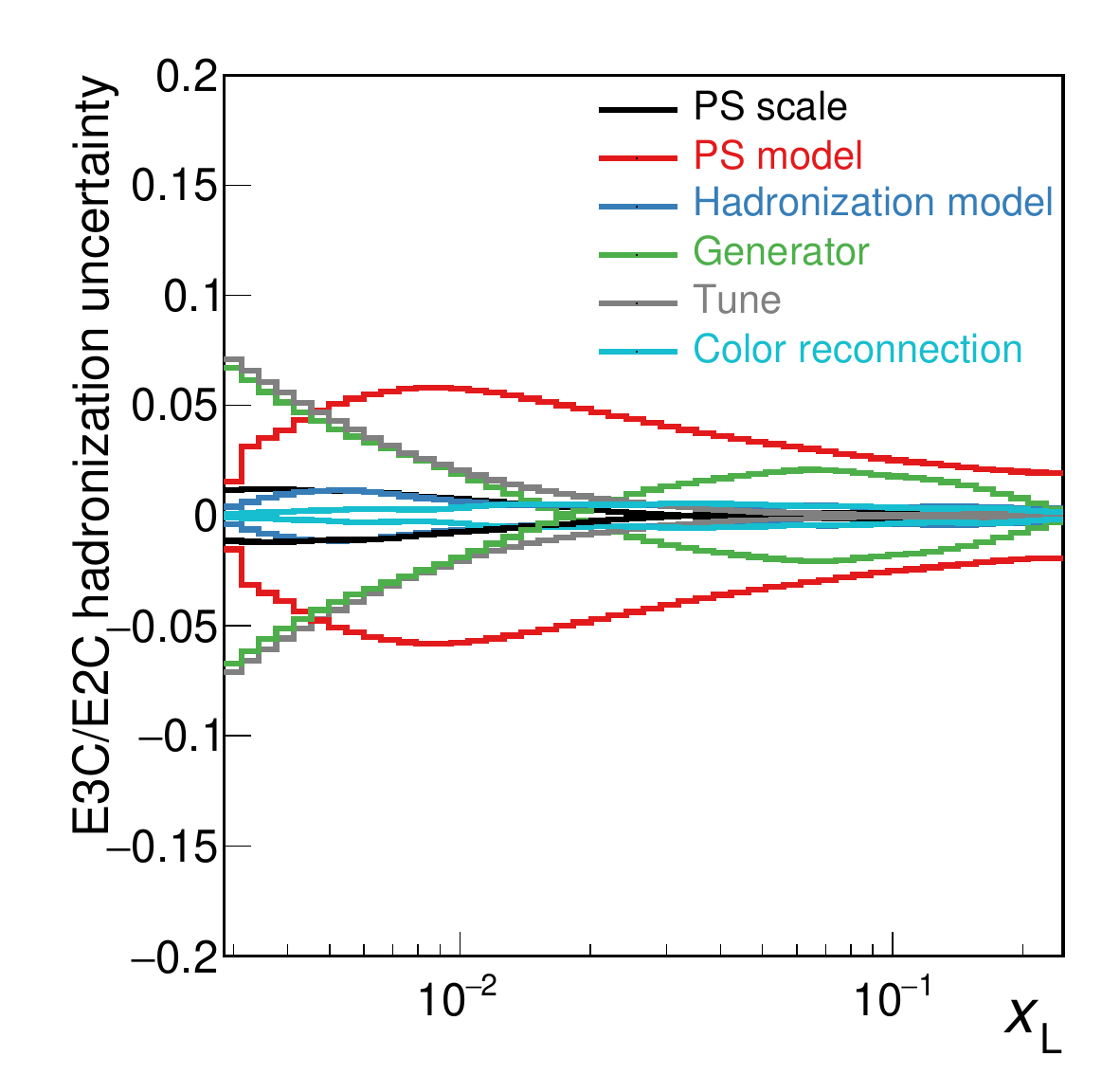}
  \caption{Hadronization uncertainty breakdown for E2C (left) and E3C/E2C (right).}
\label{fig:combined_hadronization_uncertainty}
\end{figure*}

The hadronization correction derived from various MC generators, as well as $\Omega$ from the analytical method, for E2C and E3C/E2C are shown in Fig.~\ref{fig:combined_hadronization_effects}. The correction is much more pronounced in E2C, particularly in the small \xL region. The differences between MC generators also increase towards collinear region. Using the E3C/E2C ratio largely reduces the scale of the correction, as well as the dependency on different models, which amounts to 200\% in E2C and reduces to 20\% in the E3C/E2C ratio. It is also noted that the analytical correction $\Omega$ is not close to any of the MC based corrections in E3C/E2C variable. 

The breakdown of the individual factors that affects the MC hadronization correction is extracted as described above, either from parameter variation, or from two-point comparisons between different settings. Figure~\ref{fig:combined_hadronization_uncertainty} presents the individual source of hadronization uncertainties on E2C (left) and E3C/E2C ratio (right). As expected, each single uncertainty decreases as \xL increases and gets significantly smaller switching from E2C to E3C/E2C. Among all the factors, the tune parameter and the generator differences play major roles in the most collinear region, and the PS model becomes dominant in the slightly larger \xL region. PS scale, hadronization model and CR model uncertainties are smaller in size, all within 2\% for E3C/E2C. However, we would like to point out that when it comes to \as extraction, not only the size of the uncertainty, but also the shape is important, as all the bins are treated correlated.

\section{\label{Detector}Detector effects and experimental uncertainties}
Before estimating the \as sensitivity, we assess the detector effects in realistic experimental conditions. Monte Carlo events are generated using \PYTHIA{8.306} with $p_T$-ordered PS, Lund string hadronization model and Monash tune, followed by detector simulation through Delphes~\cite{delphes}, implementing both CEPC~\cite{Chen:2017yel} and FCC-ee~\cite{FCC:2018evy} detector configurations. The CEPC detector design has an Electromagnetic Calorimeter (ECAL) with an energy resolution of approximately $16\%/\sqrt{E}$, and a Hadronic Calorimeter (HCAL) achieving an energy resolution of around $50\%/\sqrt{E}$ \cite{JIANG2020164481}. The FCC-ee detector has a similar performance in its calorimetric system, with the ECAL achieving an energy resolution of $16.5\%/\sqrt{E}$, and the HCAL delivering a slightly improved resolution of $44.2\%/\sqrt{E}$ \cite{Aleksa:2021ztd}. Both experiments employ the Particle Flow Algorithm (PFA) to optimize their detection capabilities and expect exceptional angular resolution for charged particles, reaching precisions of $\leq 0.1$ mrad for momenta above 10 GeV \cite{thecepcstudygroup2018cepc,FCC:2018evy}, which provides excellent precision for energy correlator measurements.

\begin{figure}[htb]
  \centering
    \includegraphics[width=0.48\textwidth]{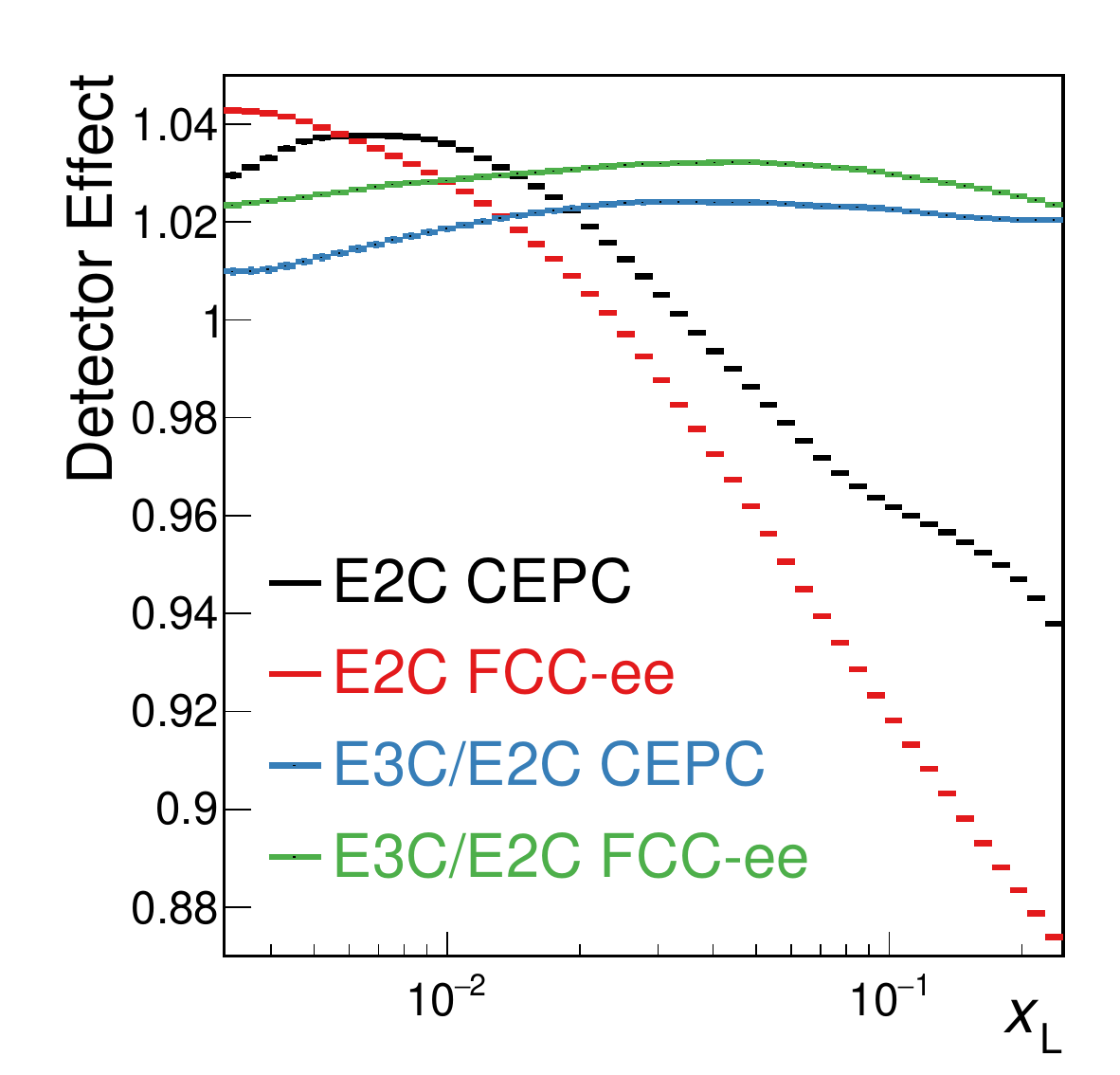}
      \caption{Detector effects on the E2C and E3C/E2C in CEPC and FCCee.}
  \label{fig:detector_effect}
\end{figure}

\begin{figure*}[htb]
  \centering
    \includegraphics[width=0.48\textwidth]{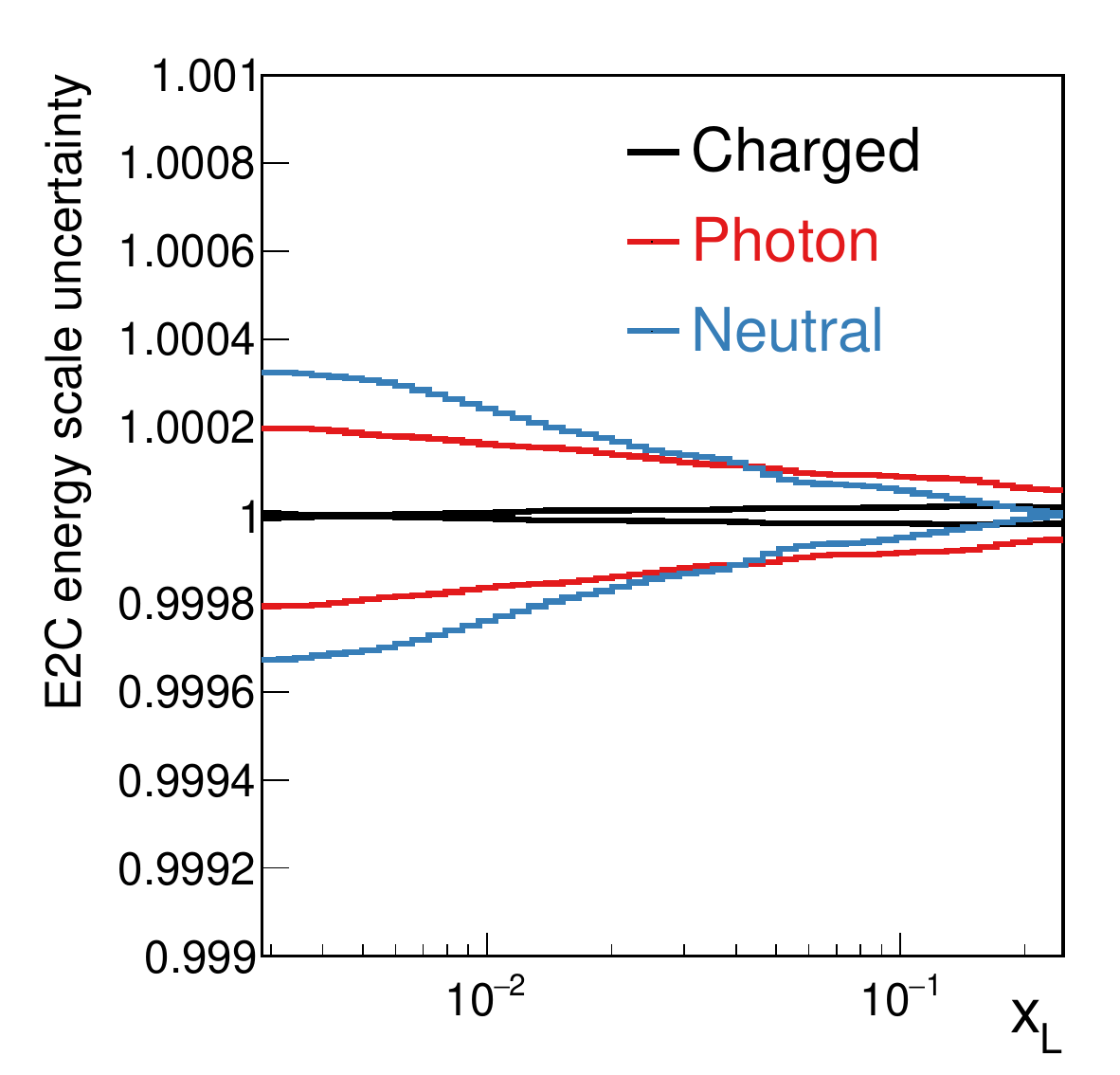}
    \includegraphics[width=0.48\textwidth]{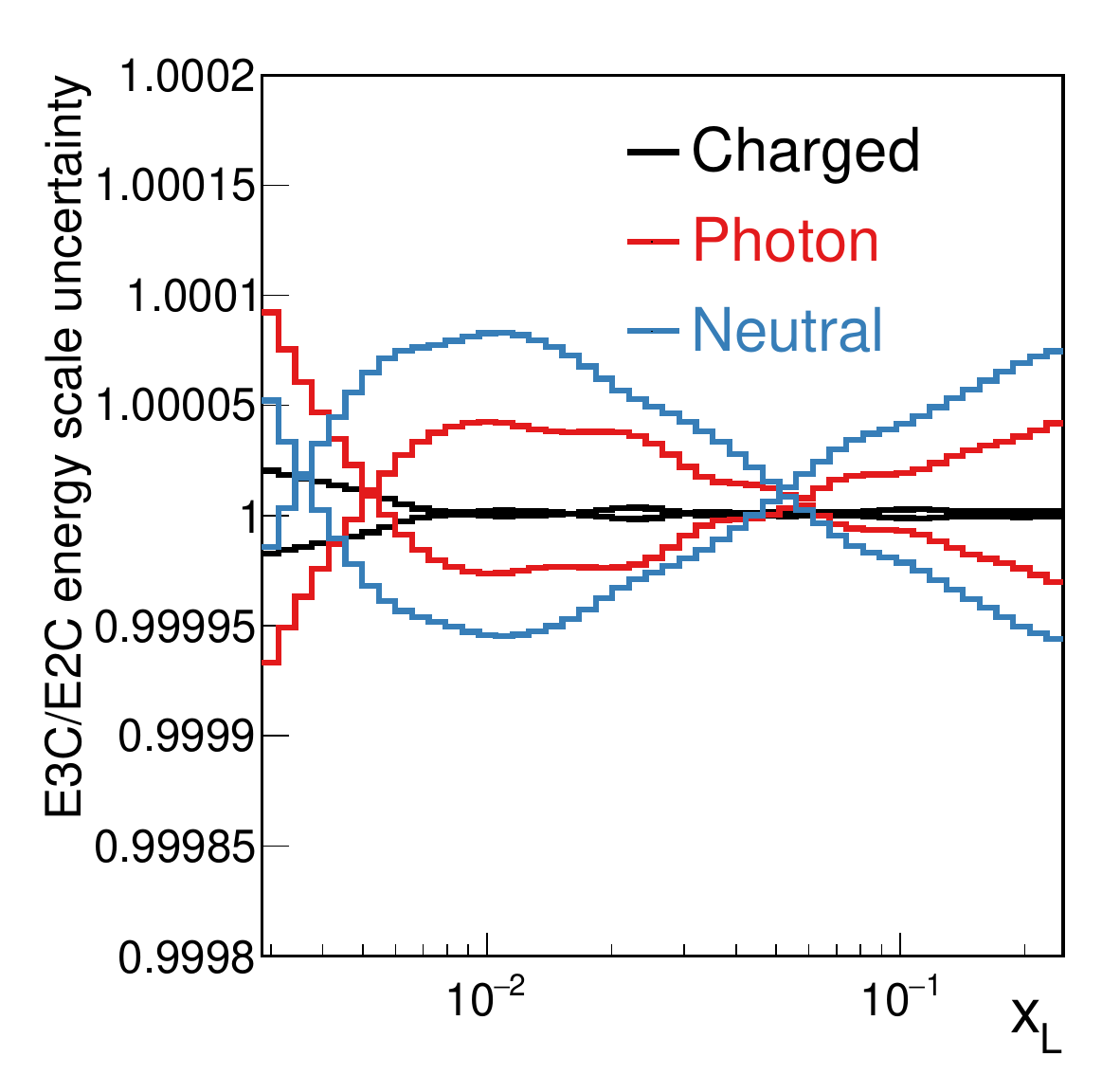}
  \caption{Particle energy scale uncertainty for E2C (left) and E3C/E2C (right).}
  \label{fig:exp_unc}
\end{figure*}

The detector effects are evaluated through the ratio of distributions at hadron level before and after the Delphes detector simulation. Figure~\ref{fig:detector_effect} illustrates the detector effects on E2C and E3C/E2C ratio for CEPC and FCCee. The two experiments exhibit similar performances, the effect is between 5--10\% for E2C depending on the \xL value. The experimental effect cancels out in the E3C/E2C ratio, both in magnitude and shape dependencies. The size is 1--3\% for both experiments, with a slightly smaller size in CEPC. As the detector performances are rather similar, we use CEPC to evaluate potential experimental uncertainties and further demonstrate the expected sensitivity of \as.

In earlier studies, CMS has extracted \as from E3C/E2C and found the dominant experimental uncertainty being the energy scale of the composite particles~\cite{CMS:2024mlf}. Therefore we focus on these sources to evaluate the overall experimental uncertainty. Depending on the particles' types, the energy scale uncertainties at CMS are 1\%, 3\% and 5\% for charged particles, photons and neutral particles, respectively. With the excellent tracker and calorimeter design of CEPC and FCCee, it is safe to assume the uncertainties could be reduced by half. We vary the energy scales of the individual particles entering the energy correlator calculation, and the impact of the uncertainties is presented in Fig.~\ref{fig:exp_unc}. For the E2C observable, the uncertainties are at the scale of 0.04\%, with neutral hadrons being the largest source of uncertainty. The effect is further reduced in E3C/E2C ratio, only up to 0.01\% level. This demonstrates the robustness of the observable against detector-related systematic uncertainties. Given that these uncertainties are several orders of magnitude smaller than the hadronization and theoretical uncertainties, we neglect their contribution in the subsequent \as extraction.

\section{\label{result}Expected sensitivity of \as}

We apply the hadronization and detector corrections to the above \NNLL parton-level predictions at $\asmz=0.118$ to obtain the pseudo data distributions of E3C/E2C. The hadronization factor is taken from \PYTHIA and the detector effects are taken from CEPC setting. The distributions are scaled to the luminosity of 40 \pbinv and compared to predictions under a set of different values of \asmz, where hadronization and detector effects are also propagated. 
The $\chi^2$ between the pseudo data and the predictions are calculated to assess the expected sensitivity of $\as$, including uncertainties in theoretical predictions discussed in Sec.~\ref{Theory} and hadronization effects. We compare three approaches in handling the hadronization uncertainty: scheme1 (S1) follows the traditional MC method, which takes the overall difference between predictions from \HERWIG and \PYTHIA generators, where all the default settings in the generators are taken; Scheme2 (S2) does a more detailed assessment by decomposing the MC based hadronization effects into six distinct sources of uncertainties as discussed in Sec.~\ref{Simulation}; Scheme3 (S3) tries to take into account the difference between the MC and analytical predictions. As there are no factorized way to evaluate the uncertainty, we take the difference between \PYTHIA and $\Omega$, and the difference between \HERWIG and $\Omega$ as two independent systematic sources. The experimental uncertainties have shown a negligible impact and are safely discarded. 

The bins in energy correlator distributions are not statistically independent. The correlation matrix among them is recorded and used in the $\chi^2$ calculation, which is defined as

\begin{equation}
\begin{aligned}
\chi^2\left(\as, \vec{\theta}\right) = & \left(\vec v_{\text{th}}\left(\vec{\theta}, \as\right)
-\vec v_{data}(\vec\theta)\right)^T V_{\text{data}}^{-1}
\\
&\left(\vec v_{\text{th}}\left(\vec{\theta}, \as\right)
-\vec v_{data}(\vec\theta)\right)
 +\sum_i \theta_i^2 .
\end{aligned}
\label{eq:chi2}
\end{equation}

Here, $\vec v_{\text{th}}$ represents the theoretical prediction of E3C/E2C distribution for various $\asmz$, and $V_{\text{data}}$ is the covariance matrix derived from pseudo data. The hadronization and theoretical uncertainties enter the $\chi^2$ as shape nuisance parameters $\vec{\theta}=\left(\theta_1, \theta_2\right)$, which means all the bins are varied correlatively under the uncertainties. Each changes the shape of E3C/E2C and constraint by a Gaussian distribution. The overall $\chi^2$ is minimized floating $\vec{\theta}$ at each value of \asmz.

\begin{figure*}[b]
  \centering 
  \includegraphics[width=0.8\textwidth]{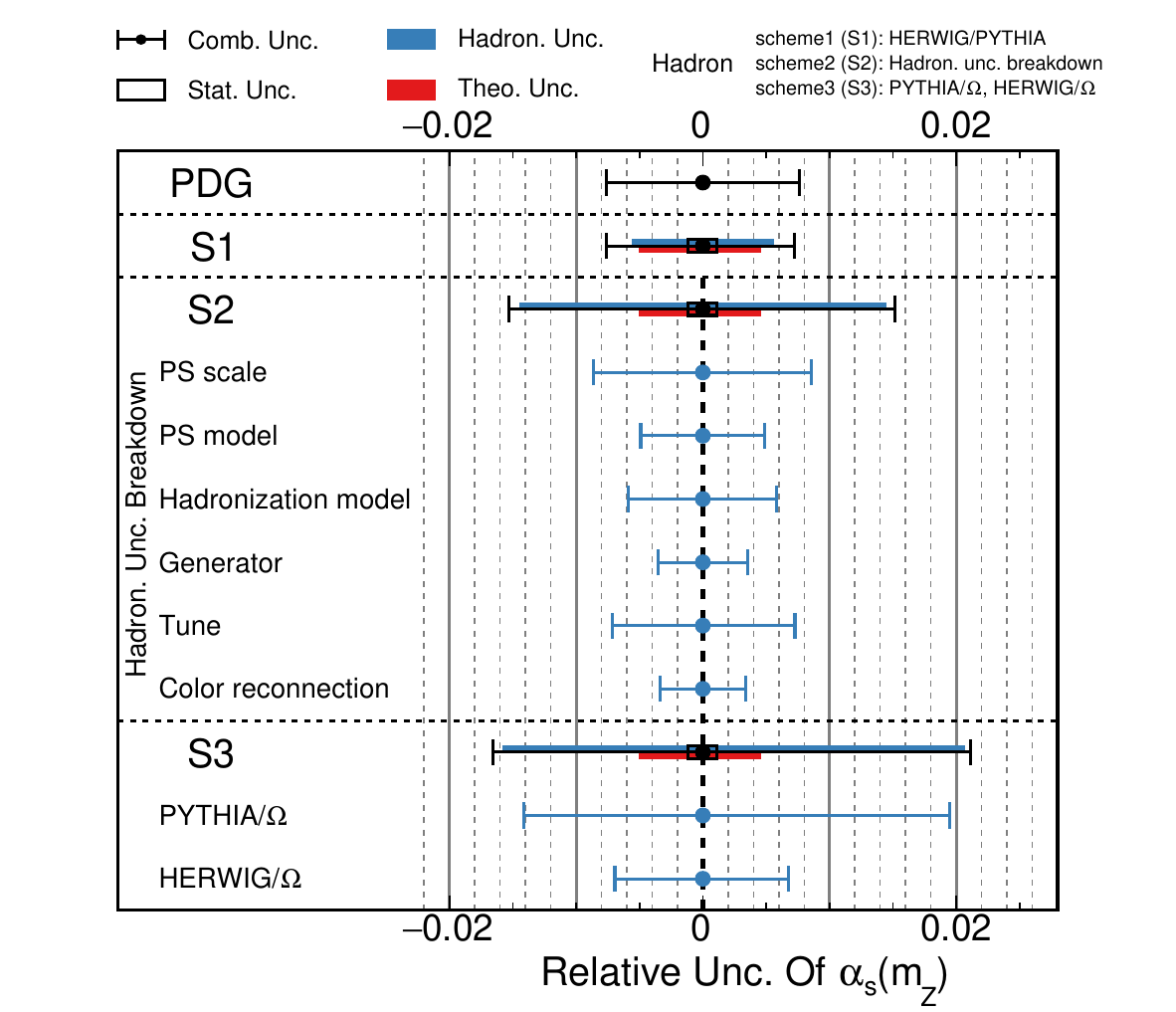} 
  \caption{The expected sensitivity to \asmz using E3C/E2C at CEPC at 40 \pbinv. The PDG 2022 world average precision for $\as$ extraction is shown for a comparison~\cite{pdg2022}. The breakdown of statistical, hadronization, and theoretical uncertainties, as well as the combined results, are shown. Three approaches for evaluating hadronization uncertainties are presented. S1 uses the traditional method based on the difference between \HERWIG and \PYTHIA predictions. S2 employs a detailed decomposition into six distinct sources and the breakdown of these components is presented. S3 takes the difference between the analytical and MC based predictions. } 
  \label{fig:chi2_result}
\end{figure*}

We extract the expected sensitivity of \asmz from the $\chi^2$ scan at an integrated luminosity of 40 \pbinv, which is comparable to the integrated luminosity used in the previous \as extraction with the E2C observable~\cite{kardosetal2018} from \epem data. The CEPC aims to collect a large dataset of $100\ \text{ab}^{-1}$ over two years at $\sqrt{s}=91.2$ GeV~\cite{cepcacceleratorstudygroup2022snowmass2021, Zhu:2022hyy}, and the FCC-ee plans to take 150 $\text{ab}^{-1}$ data in four years~\cite{FCC:2018evy}. Therefore the considered luminosity could be achievable at both experiments in less than 1 minute.

Figure~\ref{fig:chi2_result} shows the expected uncertainties on \asmz. The theoretical uncertainty contributes about 0.5\% and the statistical uncertainty is 0.1\%. The contribution of hadronization uncertainty as well as the overall expected sensitivity is shown separately in the three schemes.
In S1, the same two-point hadronization uncertainty adopted by the previous \epem E2C result is used. The hadronization uncertainty using E3C/E2C is approximately 0.6\%, yielding an overall sensitivity of 0.8\%. This is significantly better than the 2.4\% reported in the previous E2C result, illustrating the improved control over uncertainties offered by the E3C/E2C ratio observable.
In S2, a more comprehensive decomposition of the hadronization effects is performed, and a larger hadronization uncertainty of 1.5\% is obtained. The contribution of each single source is also shown in Fig.~\ref{fig:chi2_result}. All of them show sizable impacts on the \as determination, among which the PS scale and tune settings contribute the most. The difference seen in S1 and S2 suggests that a simple two-point variation may underestimate the MC hadronization uncertainty. 
In S3, we see a large difference between analytical prediction $\Omega$ and \PYTHIA, and the hadronization uncertainty reaches 2.1\%.

With 40 \pbinv data, we see that the uncertainties of the theoretical prediction and the hadronization corrections are already dominated over statistical uncertainty. In particular, different treatment of the hadronization uncertainty could change the result significantly. Therefore extra efforts would be needed to understand these important effects and reduce the size of the relevant uncertainty in the future, to better exploit the large dataset going to be collected in \epem colliders.

\section{\label{Conclusion}Conclusion}

In this study, we evaluate the sensitivity of \as extraction at future \epem colliders by using E3C/E2C ratio observable in the collinear region, which has not been explored by previous event-shape based methods. This observable demonstrates substantial advantages, including minimal dependence on detector response and reduced hadronization effects. With \NNLL precision calculation, several schemes for evaluating hadronization uncertainties are considered. In the traditional approach that only considers differences between Monte Carlo models like \PYTHIA and \HERWIG, the expected relative sensitivity could reach approximately 0.8\% under a luminosity of 40 \pbinv. However, detailed decomposition into sources including parton shower scale and model, hadronization model, tune, color reconnection model and generator implementations, a more conservative sensitivity of 1.5\% is expected. If the difference between analytical and MC based methods are considered, the sensitivity degrades to 2.1\%. The study highlights the importance of understanding hadronization effects beyond simple model comparisons. The theoretical uncertainty, along with the hadronization uncertainty, have been shown to be the limiting factors in extracting \as through energy correlators in future high luminosity \epem colliders and a better understanding of them is in need.

\section*{Acknowledgements}
We thank Huaxing Zhu for the useful discussions. The work is supported by National Natural Science Foundation of China (NSFC) under the Grant No. 12322504 and the
Innovative Scientific Program of the Institute of High Energy Physics.

\bibliography{references}

\end{document}